\documentclass[doublespacing,twocolumn]{elsart}

\usepackage{epsfig}

\usepackage{amssymb}

\begin{document}

\begin{frontmatter}



\title{Superconductivity, d Charge-Density Wave and Electronic
Raman Scattering in High-T$_c$ Superconductors}


\author{R. Zeyher$^a$ and A. Greco$^b$}

\address[a]{Max-Planck-Institut f\"ur Festk\"orperforschung,
70569 Stuttgart. Germany}
\address[b]{Departamento de F\'{\i}sica, Av. Pellegrini 250,
2000-Rosario, Argentina}

\begin{abstract}
The competition of superconductivity and a d charge-density wave (CDW)
is studied in the $t-J$ model as a function of temperature at large $N$ where
$N$ is the number of spin components. Applying the theory to electronic
Raman scattering the temperature dependence of the $B_{1g}$ and the
$A_{1g}$ spectra are discussed for a slightly underdoped case.

% Text of abstract
\end{abstract}

\end{frontmatter}

% main text
%\section{Introduction}
\vspace{-0.2cm}
Generalizing the $t-J$ model from two to N spin components
it can rigorously be shown that its phase diagram 
exhibits a quantum critical point at large N
at a critical doping $\delta = \delta_0$. It separates a superconducting
ground state with a d-wave order parameter $\Delta({\bf k})$ for 
$\delta > \delta_0$ from a state with two competing d-wave order parameters,
namely $\Delta$ and a d CDW wave order parameter 
$\Phi({\bf k})$\cite{Cappelluti,Chakravarty,Zeyher}. Below we will
present results on the temperature dependence of the two
order parameters and the resulting electronic Raman spectra.

\vspace{-0.4cm}
Fig. 1 shows the temperature dependence of the maximum amplitude
of the two order parameters, $\Delta$ and $\Phi$, 
\begin{figure}[h]
\vspace*{-0.5cm}
\centerline{
      \epsfysize=6cm
      \epsfxsize=7cm
      \epsffile{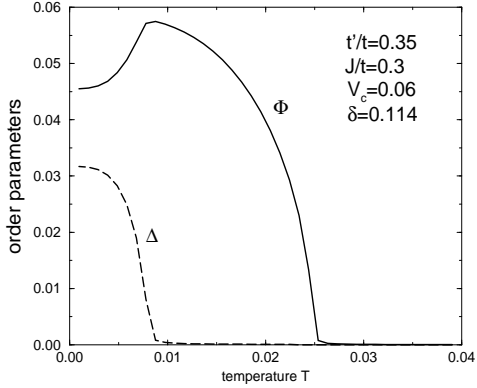}}
\vspace*{-0.8cm}
\label{figRio1}
\caption{$\Phi$ and $\Delta$ as a function of
temperature for the doping $\delta = 0.114$.}
\end{figure}
using the
nearest hopping amplitude $t$ as the energy unit. The doping $\delta = 0.114$ 
corresponds to a slightly underdoped case because $\delta_0 \sim 0.15$ 
determines
essentially optimal doping. $V_c$ is the Coulomb repulsion between 
nearest neighbors. $\Phi$ sets in at
high temperatures in a mean-field like square root fashion, whereas
$\Delta$ exhibits a smooth onset. The curves demonstrate
that the increase of $\Delta$ is accompanied by a decrease in $\Phi$
at low temperatures reflecting the competition between the two
order parameters.

\begin{figure}[h]
\vspace*{-0.5cm}
\centerline{
      \epsfysize=6cm
      \epsfxsize=7cm
      \epsffile{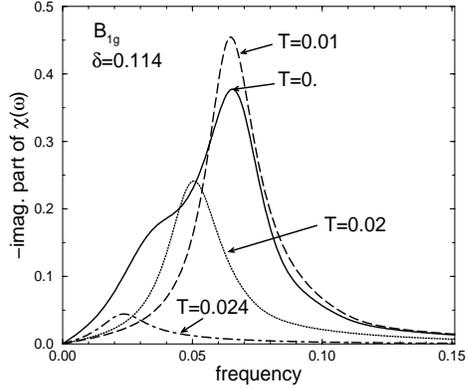}}
\vspace*{-0.8cm}
\label{figRio2}
\caption{Electronic $B_{1g}$ Raman spectra for 
$\delta =0.114$, $t'/t=-0.35$, $J/t=0.3$, and different
temperatures.}
\end{figure}

\vspace{-0.3cm}
Fig. 2 shows $B_{1g}$ spectra for the same doping at different
temperatures. The main peak at $T=0$ lies well below
the excitations of free quasiparticle across
the gap and corresponds to a collective excitation which may be viewed
either as an exciton state of the superconductor or as the amplitude
mode of the d CDW order parameter. The shoulder 
in the $T=0$ curve is caused by transitions over the
superconducting gap and may have been observed in underdoped 
$YBa_2Cu_3O_{7-\delta}$\cite{Slakey}. This sideband decreases with 
decreasing $\Delta$
and vanishes at $T=0.01$. The main peak does not
move within this temperature interval which means that it reflects the
behavior of the total gap. With increasing temperature the peak moves
to smaller frequencies and looses rapidly in intensity.
\begin{figure}[h]
\vspace*{-0.5cm}
\centerline{
      \epsfysize=6cm
      \epsfxsize=7cm
      \epsffile{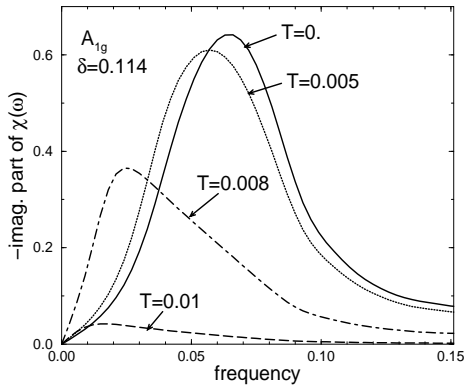}}
\vspace*{-0.8cm}
\label{figRio3}
\caption{Electronic $A_{1g}$ Raman spectra for 
$\delta =0.114$, $t'/t=-0.35$, $J/t=0.3$, and different
temperatures.}
\end{figure}

\vspace{-0.3cm}
Density fluctuations may couple to $\Delta$ via their modulation of
the density of states at the Fermi energy. Using this coupling
Fig. 3 shows calculated $A_{1g}$ spectra for different temperatures.
The broad peaks are again collective in nature and describe ampliude
fluctuations of $\Delta$. Their maxima are appximately given by 
$2\Delta$. The peak positions and their intensities decrease
rapidly with increasing temperature with the onset of $\Delta$ at
around $T\sim 0.01$ setting the scale. The usual density coupling
may also contribute to the $A_{1g}$ spectrum, especially, if the
Fermi surface has several sheets. The corresponding temperature dependence
is similar to that of the $B_{1g}$ spectrum in Fig. 2. Different
temperature dependencies of spectra thus could discriminate between
different coupling mechanisms.
\label{}
% \appendix
% \section{}
% \label{}

\end{document}